\documentstyle[12pt]{article}

\newcommand{\nc}{\newcommand}
\nc{\half}{\mbox{\small$\frac{1}{2}$}}
\nc{\quart}{\mbox{\small$\frac{1}{4}$}}
\nc{\eighth}{\mbox{\small$\frac{1}{8}$}}
\nc{\threehalf}{\mbox{\small$\frac{3}{2}$}}
\nc{\oneon}{\mbox{\small$\frac{1}{N}$}}
\nc{\be}{\begin{equation}}
\nc{\ee}{\end{equation}}
\nc{\bea}{\begin{eqnarray}}
\nc{\eea}{\end{eqnarray}}
\nc{\mc}{\multicolumn}
\title{ \vbox{\vspace{-20cm}}
An Improved Estimator for the Correlation Function
of 2D Nonlinear Sigma Models \\
}
\author{\vbox{\vspace{10mm}}
   {\bf Martin Hasenbusch }
   \\[9mm]
      CERN, Theory Division,\\[-1mm]
      CH-1211 Gen\`eve 23, Switzerland $^1$\\[2mm]
}
\date{}
\begin{document}

\maketitle

\vspace*{-11.0cm}
 \hfill CERN-TH.7375/94

\vspace{-0.5mm}

 \hfill August 1994

\vspace*{+10.0cm}

\begin{abstract} \normalsize 
\vspace{2mm}

I present a new improved estimator for the correlation function
of 2D nonlinear sigma models. Numerical tests for the 2D $XY$ model
and the 2D $O(3)$-invariant vector model were performed.
 For small physical volume,
i.e. a lattice size small compared to the bulk correlation length, a
reduction of the statistical error of the finite system
correlation length by a factor of up to 30
compared to the cluster-improved estimator was observed.  This improvement
allows for a very accurate  determination of the running coupling
proposed by M. L\"uscher et al.\ for 2D $O(N)$-invariant vector models.

\end{abstract} \vfill
\vspace{1.5cm}
 CERN-TH.7375/94  \newline
\vspace{-0.5mm}
 August 1994
\vspace{0.5cm} \newline
$^1$ Address after September 30, 1994:  DAMTP, Silver Street,
Cambridge, CB3 9EW, England

\thispagestyle{empty}

\newpage

\section{Introduction}
Almost twenty years ago Nightingale \cite{Night}
 pointed out that the correlation
length of a finite system can be used to trace the renormalization group
flow of a statistical model. He demonstrated the validity of his approach
at the example of the 2D Ising model using its exact solution
\cite{Onsager}.

L\"uscher computed the mass gap (the inverse correlation length) for the 2D
$O(3)$ vector model on finite lattices using perturbation
 theory \cite{Luescher1}.
In ref. \cite{Luescher2} L\"uscher, Weisz and Wolff used the mass gap of finite
systems
to define a running coupling for 2D asymptotically free models. The aim of
their study was to match the perturbative, i.e. small distance, regime
with the nonperturbative
large distance regime, using MC simulations for large distances
and in the cross-over region.
The high numerical accuracy that  they reached
 for the step scaling function of this particular running coupling
of the 2D $O(3)$-invariant vector
model was based
on the use of the efficient cluster algorithm \cite{single}
combined with a variance reducing
cluster-improved estimator \cite{Wolffo3}. However one has to
note that up to now this powerful
simulation technique only applies to a rather restricted class of models.
 I will discuss an alternative improved estimator that
does not rely on the cluster algorithm and hence can  be applied
to a larger class of symmetry groups.

 The concept of
improved estimators is rather general and by no means
restricted to the
cluster algorithm. The basic idea of improved estimators
is to integrate analytically or by standard numerical methods those
degrees of freedom that are most important for the fluctuations of the
observable in question.
For local fluctuations this can be achieved easily. Parisi et al.
\cite{Parisi} proposed to improve the estimator of the Wilson-loop
correlation-function by integrating over the link-variables on the
Wilson-loops.

I will show that such ideas also apply
to large range correlation functions of certain models.
The estimator is based on an embedding of a one-dimensional model.
The degrees of freedom of this embedded model are collective transformations
 of all spins
in a given time-slice.

This paper is organized as follows. In section 2 I define the new
improved estimator for two-dimensional $O(N)$
invariant vector models.
 In section 3 I summarize theoretical predictions for the running coupling
of 2D $O(N)$-invariant vector models.
 In section
4 I present numerical results for the 2D $O(2)$-invariant vector ($XY$)
 model and the 2D $O(3)$-invariant
vector model. The conclusions are given in section 5.
\newpage
\section{The improved estimator}
First let us discuss the
strategy applied for finding an improved estimator for an observable $A$ of a
statistical model with field variables $\phi$. The expectation value of
$A$ is given by
\be
\langle A \rangle = \frac {\int D\phi \exp(-S(\phi)) A(\phi)}
                       {\int D\phi \exp(-S(\phi)) } \; .
\ee
We split the set of field variables into two parts
 $\phi = (\bar{\phi},\Phi)$, where
$\Phi$ is chosen such that the integrals over $\Phi$ can be performed exactly
for any fixed $\bar{\phi}$.
Let us define the improved estimator by
\be
A_{imp}(\phi) = A_{imp}(\bar{\phi}) = \langle A \rangle_{cond} =
 \frac {\int D\Phi \exp(-S(\bar{\phi},\Phi)) A(\bar{\phi},\Phi)}
                       {\int D\Phi \exp(-S(\bar{\phi},\Phi)) } \; ,
\ee
where $ \langle ... \rangle_{cond}$ denotes the conditional, i.e. $\bar{\phi}$
dependent expectation value.  It is easy to check that
  $\langle A_{imp} \rangle = \langle A \rangle $
and also that the variance of
the improved estimator $A_{imp} $ is reduced compared to that of $A$
\be
\langle A^2 \rangle -\langle A \rangle^2 \ge
 \langle A_{imp}^2 \rangle -\langle A_{imp} \rangle^2 \; .
\ee
Of course, the remaining non-trivial question is to identify, for a given
model and observable, the $\Phi$ such that the integration over $\Phi$ can
be performed with reasonable effort and leads to a considerable reduction of
the variance.

Now let me explain the improved correlation function estimator
for the $O(N)$-invariant
vector models.
The action is given by

\be
\label{action}
 S=-\beta \sum_{x,t} \sum_{\alpha=1}^{N}
 \left( s_{\alpha}(x,t) s_{\alpha}(x+1,t)
       +s_{\alpha}(x,t) s_{\alpha}(x,t+1)   \right)
\ee
where $s_{\alpha}(x,t)$ are the components of a real $N$-component unit vector.
The lattice has extension $T$ in the time ($t$-) direction and
extension $L$ in
spatial ($x$-) direction.
For simplicity we set the lattice-spacing $a$ equal to $1$ in this section.
In time direction we employ open boundary
conditions while in $x$-direction periodic boundary conditions are employed,
i.e. identify $x=L+1$ with $x=1$ and $x=0$ with $x=L$.
We choose  left-multiplications of
 all  spins in a time-slice $t$ with the same
 $O(N)$-matrix $R(t)$ as degrees of freedom of the 1D embedded
model. The improved estimator is then constructed by integrating over the
$R(t)$ exactly.
 The conditional, i.e. $s_{\alpha}$ dependent,
 action of the embedded models is given by
\be
 S_{cond}(R) =-\beta \sum_{x,t} \sum_{\alpha=1}^{N}
\left( \sum_{\delta=1}^{N}   R_{\alpha,\delta}(t) s_{\delta}(x,t) \right)
\left( \sum_{\gamma=1}^{N}  R_{\alpha,\gamma}(t+1) s_{\gamma}(x,t+1) \right)
\ee
Note that contributions from spatial bonds stay constant under the
transformation, and can hence be ignored.
The conditional partition function is given by
\be
Z_{cond} = \int DR  \exp(-S_{cond}(R)) \; ,
\ee
where $DR = \prod_t dR(t)$ and $dR$ denotes
 the Haar measure of the $O(N)$ group.
Next let us introduce relative rotations $X(t,t+1)$ of two neighbouring
time-slices by
\be
X(t,t+1) = R^{-1}(t) R(t+1)
\ee
All $R(t)$ can be expressed in terms of $R(1)$ and the $X(t,t+1)$.
\be
R(t)= R(1) X(1,2) ...  X(t-2,t-1) X(t-1,t)
\ee
Now it becomes clear why free boundary conditions in time direction
are used.
Periodic boundary conditions in time direction
would require that the product of all $X(t,t+1)$
on the lattice is equal to the unit matrix.
In terms of the $X$ the action of the conditional model is given by
\be
 S_{cond}(X) =-\sum_{t} \sum_{\alpha=1}^{N}  \sum_{\gamma=1}^{N}
 X_{\alpha,\gamma}(t,t+1) Q_{\alpha,\gamma}(t,t+1)
\ee
where $Q$ is given by
\be
 Q_{\alpha,\gamma}(t,t+1) = \beta \sum_x s_{\alpha}(x,t) s_{\gamma}(x,t+1) \;.
\ee
The properties of the Haar measure  allow also to rewrite the measure in terms
of the relative rotations $X$
\be
Z_{cond} =\int dR(1) \int DX  \exp(-S_{cond}(X)) \; ,
\ee
where the integration over $R(1)$ just gives a constant factor that hence
can be ignored.
The conditional partition function
factorizes
in two-slice partition functions
\be
Z_{cond} = \prod_{t=1}^{t=T-1} z(t,t+1) \;
\ee
with
\be
 z(t,t+1) = \int dX  \exp \left(-\sum_{\alpha=1}^{N}  \sum_{\gamma=1}^{N}
 X_{\alpha,\gamma}(t,t+1) Q_{\alpha,\gamma}(t,t+1)  \right) \; ;
\ee
where $dX$ denotes the Haar measure of the $O(N)$ group.

Next let us turn to the measurement of time-slice
correlation functions. The time-slice correlation function is defined by
\be
\label{con}
G(\tau) = \sum_{\alpha}
 S_{\alpha}(t)  S_{\alpha}(t+\tau)  \; ,
\ee
where
\be
S_{\alpha}(t) = \sum_x s_{\alpha}(x,t) \; .
\ee
The improved estimator of the  time-slice correlation function is given by
\be
G_{imp}(\tau) = \langle G(\tau) \rangle_{cond} =
 \sum_{\alpha} \Bigl\langle \sum_{\delta} R_{\alpha,\delta}(t) S_{\delta}(t)
\sum_{\gamma} R_{\alpha,\gamma}(t+\tau) S_{\gamma}(t+\tau)
 \Bigr\rangle_{cond}  \; ,
\ee
where
\be
\langle A(R) \rangle_{cond} = \frac{\int DR \exp(-S_{cond}(R)) A(R)}{Z_{cond}}
\ee
is the conditional expectation value of an observable $A(R)$.
Again we can re-express things in terms of relative rotations
\bea
\label{imp}
G_{imp}(\tau) &=&
 \sum_{\alpha} \sum_{\gamma} \Bigl\langle S_{\alpha}(t)
 X_{\alpha,\gamma}(t,t+\tau) S_{\gamma}(t+\tau)
 \Bigr\rangle_{cond} \\ \nonumber
               &=&
 \sum_{\alpha} \sum_{\gamma} S_{\alpha}(t) \Bigl\langle
 X_{\alpha,\gamma}(t,t+\tau)
                    \Bigr\rangle_{cond} S_{\gamma}(t+\tau) \; ,
\eea
where $X(t,t+\tau) = R^{-1}(t) R(t+\tau) $. Using the laws of
matrix calculus and the factorization of the conditional partition function
we obtain
\be
\Bigl\langle X(t,t+\tau) \Bigr\rangle_{cond}
=\Bigl\langle X(t,t+1) \Bigr\rangle_{cond} ...
\Bigl\langle X(t+\tau-1,t+\tau) \Bigr\rangle_{cond}\;.
\ee
Hence the evaluation of the conditional expectation values is reduced to the
evaluation of the low dimensional integral
\be
\label{integ}
\Bigl\langle X(t,t+1) \Bigr\rangle_{cond} =
\frac{\int dX \exp(\sum_{\alpha,\gamma}
   X_{\alpha,\gamma} Q_{\alpha,\gamma}(t,t+1) ) X}
     {\int dX \exp(\sum_{\alpha,\gamma}
   X_{\alpha,\gamma} Q_{\alpha,\gamma}(t,t+1) ) }
\ee
\subsection{Performing the integrals for N=2 and 3. }
For $N=2$ this integral can be solved in closed form.
One obtains
\be
 X_{11} = X_{22} = \frac{I_1(\kappa)}{I_0(\kappa)} \;
\frac{Q_{11}+Q_{22}}{\kappa}
\ee
and
\be
 X_{21} = - X_{12} = \frac{I_1(\kappa)}{I_0(\kappa)} \;
\frac{Q_{21}-Q_{12}}{\kappa} \; ,
\ee
where
\be
\kappa=
\sqrt{(Q_{11}+Q_{22})^2 + (Q_{21}-Q_{12})^2}
\ee
and $I_0$ and $I_1$ are modified Bessel-functions.

For $N=3$ I could not find an exact solution for the integrals.
For practical reasons I restricted the integration to the $SO(3)$ subgroup
of $O(3)$.

 Using the property
\be
\int dX f(X) = \int dX f(U X V)
\ee
of the Haar measure, where $U$ and $V$ are elements of the group manifold,
the integrand
\be
\int dX \exp(\mbox{Tr} \; X Q^T) = \int dX \exp(\mbox{Tr} \; U X V Q^T ) =
\int dX \exp(\mbox{Tr} \; X V Q^T U )
\ee
 can be transformed such that
the source $\tilde Q =  V Q^T U$ becomes diagonal.

Now I parametrized the    $SO(3)$  using the Euler-angles. One of the 3
integrations can be performed analytically. The remaining 2D integral
I solved numerically.

If all three sums $q_1=\tilde Q_{11}+\tilde Q_{22}$,
$q_2=\tilde Q_{11}+ \tilde Q_{33}$ and
$q_3=\tilde Q_{22}+ \tilde Q_{33}$
are positive the integrand takes its maximal value
for $X$ equals the unit matrix. I performed a saddle-point
approximation around this maximum up to  order $q^{-8}$.
Comparing with the numerical solution of the integral I checked that
for $\mbox{min}(q_1,q_2,q_3) > 13$ the difference between the truncated series
and the exact result for the components of $\langle X \rangle_{cond}$ is
  smaller than
$5 \times 10^{-8}$.

In the remaining cases I employed the numerical integration described above.

\section{Theoretical Predictions}

\subsection{The 2D $O(N)$-invariant vector model with $N>2$}
The 2D $O(N)$-invariant vector models can formally be defined by the continuum
action
\be
S = - \frac{1}{2 g^2} \int d^2x \;\partial_{\mu} s(x) \partial_{\mu} s(x) ,
\ee
where $s$ is a real $N$-component unit vector and $g^2$ is the coupling
constant.
 The corresponding Wilson lattice action I have already given in  eq.
(\ref{action}).
The 2D $O(N)$-invariant vector models with $N>2$ are believed to be
asymptotically free \cite{polya}.
  This means that the model has a finite correlation
length
for all finite $\beta$ and the coupling $g^2$ vanishes with decreasing
length scale.
The running coupling
\be
\bar{g}^2 = \frac{2 L}{(N-1)\xi(L)}
\ee
introduced in ref.  \cite{Luescher2} is chosen such that $g^2$ and $\bar{g}^2$
coincide at 1-loop order of perturbation theory.
The flow of $\bar{g}^2$ is given
by the Callan-Symanzik $\beta$-function
\be
\label{beta}
\beta(\bar{g}^2) = - L \frac{\partial \bar{g}^2}{\partial L} \; ,
\ee
where $\beta(u)$ is given for small $u$ by
\be
\beta(u) = -u^2 \sum_{l=0}^{\infty} b_l u^l.
\ee
The first three coefficients of this expansion are know from
perturbation theory \cite{Luescher2}
\bea
b_0 &=& (N-2)/(2\pi) \\ \nonumber
b_1 &=& (N-2)/(2\pi)^2 \\ \nonumber
b_2 &=& (N-1)(N-2)/(2\pi)^3  \; .
\eea
Of course on the lattice only finite rescalings  are allowed. Therefore
in ref. \cite{Luescher2} a continuum step-scaling function
 $\sigma(s,u)$ is introduced,
where $s=L'/L$ is the scaling factor.
  In addition the step-scaling function
on the lattice $ \Sigma(s,u,a/L)$ is affected by finite lattice-spacing
artefacts. The continuum step scaling function is recovered in the limit of
vanishing $a$.
\be
\lim_{a \rightarrow 0} \Sigma(s,u,a/L) = \sigma(s,u)
\ee
{}From the analysis of the cutoff dependence of Feynman diagrams \cite{szym}
one knows \cite{Luescher2} that up to logarithmic corrections  the leading
finite lattice-spacing artefacts are $O(a^2)$.
The dependence of the relative error
\be
\delta(u,a/L) = \frac{\Sigma(2,u,a/L)-\sigma(2,u)}{\sigma(2,u)}
              =\delta_1(u,a/L) u + \delta_2(u,a/L) u^2 ...
\ee
can be computed in perturbation theory.
Some of the results given in ref. \cite{Luescher2} for $N=3$ are summarized in
table \ref{Luetab}.

\subsection{The 2D $XY$ model}
In contrast to the 2D $O(N)$-invariant
vector models with $N>2$ the 2D $XY$ model
undergoes a phase transition at a finite $\beta$.
The most accurate value for the critical point is $\beta_c = 1.1197(5)$
\cite{matching}.
The phase transition of the 2D $XY$ model is described by the
Kosterlitz-Thouless (KT) theory \cite{KT}.

For $\beta < \beta_c$ the correlation length $\xi_{bulk}$ is finite and
as $\beta \rightarrow \beta_c$ it diverges like
\be
 \xi_{bulk} \propto \exp(C (\beta_c - \beta)^{-1/2})
\ee
where $C$ and $\beta_c$ are model dependent.
The finite $\xi_{bulk}$  implies that $\bar{g}^2$
increases linearly with $L$ in the limit of large  $L$.

For $\beta > \beta_c $ the theory is massless,
 i.e., the correlation function decays algebraically in the infinite volume.
The large distance behaviour in the massless phase is described by an
effective
Gaussian theory
\be
S_{eff} =  \frac{1}{2} \beta_{eff}  \int d^2x
  (\nabla \phi(x))^2 \; ,
\ee
where $\phi$ is the angle of the spin. The critical point
is characterized by $\beta_{eff} = 2/\pi$.
  For the Gaussian action one obtains
$\bar{g}^2 = 1/\beta_{eff}$ independent of $L$.

The improvement  of course depends on the distance $\tau$ that is considered.
In general the statistical error will increase exponentially with $\tau$.
As an example I studied the correlation function $G(\tau)$ and the effective
correlation length $\xi_{eff}(\tau)$  at $\beta=1.8932$ and $L=64$
 in more detail.

In figure 1 the relative
statistical error of the correlation function is plotted as a
 function of the distance $\tau$ for the three estimators.
The error of the
new improved estimator is smaller than that of the other estimators for all
distances considered and stays almost constant in the studied region.
In figure 2 the relative statistical
error of $\xi_{eff}(\tau)$ is plotted. As for the
correlation function itself the error of the improved estimator is strictly
smaller than that of the other two estimators and stays almost constant with
increasing $\tau$,  while the error of the other two estimators
 rapidly increases
with increasing distance $\tau$.

Next I tried to reproduce two of the results for the step-scaling function
\be
\sigma(2,0.7383) = 0.8166(9)
\ee
and
\be
 \sigma(2,1.0595) = 1.2641(20)
\ee
given in ref. \cite{Luescher2}.
Therefore I determined  $\Sigma(2,0.7383,a/L)$ and  $\Sigma(2,1.0595,a/L)$
for various $L/a$ in the range
$L/a=6$ up to $L/a=32$.
First one has to find for each $L/a$ the $\beta$ at which
 $\bar{g}^2=0.7383$  or $\bar{g}^2=1.0595$
is reached. For $L/a \le 16$ I took the $\beta$'s found in ref.
\cite{Luescher2} as first guess. In order to
correct the small deviations from the wanted value of
$\bar{g}^2$ I determined the slope of
$\bar{g}^2(\beta)$ from runs with smaller statistics (10000 measurements)
at $\beta$-values
slightly smaller and larger.
For the $L$ not studied in ref. \cite{Luescher2} I started with a guess
obtained from extrapolating the data for smaller $L$.
 Then I computed a second
guess from two  simulations with 10000 measurements in the neighbourhood
of the first guess.  The result of the simulation with 100000 measurements
at this second guess for $\beta$ I give in table 1. The final correction
is then done using the result for the slope of $\bar{g}^2(\beta)$.

Next I simulated  on a lattice of size $L'/a=s L/a$ with $s=2$
at the $\beta$-values obtained above. The results of these
simulations are also given in table 1.

 The resulting $\bar{g}^2$ gives the finite lattice-spacing
 approximation $\Sigma(s,u,L/a)$ of
 $\sigma(s,u)$.
 In order to obtain the continuum result the limit
$a \rightarrow \infty$ has to be taken. For  $\Sigma(s,1.0595,L/a)$ the
finite lattice-spacing artefacts even have a different sign than predicted
by perturbation theory \cite{Luescher2}. Nevertheless I assume,
following ref. \cite{Luescher2} that
 the finite
lattice spacing artefacts vanish  approximately
like $(a/L)^2$.
Hence one might improve the convergence  towards $\sigma(s,u)$
with decreasing $a/L$
by fitting  the data for $\Sigma(s,u,L/a)$
 according to an ansatz
\be
\Sigma(2,u,a/L) = \sigma(2,u) + c_2 \; (a/L)^2 + c_4 \; (a/L)^4 ... .
\ee
Results for such fits are given in table 3. The statistical error of the
$\beta$'s is taken into account in the statistical error of $\Sigma(2,u,a/L)$.
Even when including all data points the $\chi^2/dof$ is small.
The  result for $ \sigma(2,1.0595)$
stays stable when data-points  with small $L$
are discarded; also $c_2$ stays stable.
My final estimate $\sigma(2,1.0595) = 1.2639(10)$, taken from the fit where the
two smallest $L/a$-values are discarded,  is consistent with that of
ref. \cite{Luescher2}, while the statistical error is reduced by more than
a factor of 2.   Including larger lattice sizes also reduces the danger
of systematical errors.
A similar analysis leads to  $\sigma(2,0.7383) = 0.8156(4)$ which is again
consistent with ref. \cite{Luescher2}.

Next I simulated the model at $\beta=5.0$ on lattices of size
$L=6,8,12,16,24,32$ and 48. The statistical error of the correlation length
determined from the
conventional and the cluster estimator were approximately the same. But
 with the new estimator the statistical error
 was reduced by a factor of about 30
compared to the other two estimators.
The results for the running coupling are summarized in table 4.
For comparison I computed $\bar{g}^2$
  to 1, 2 and 3-loop order perturbation theory using the result
for L=48 as start value for the integration of eq. \ref{beta}.
The  numerical results
for $L=12,16,24$ and 32 are reproduced within two standard deviations. The
larger deviations for $L=6$ and 8 are well described by corrections to scaling
computed in perturbation theory
 \cite{Luescher2}.

Finally I also checked whether the new estimator also provides  an improvement
for the measurement of the bulk correlation length $\xi_{bulk}$. Therefore
I simulated the model at $\beta=1.5$.  The correlation
length is $\xi_{bulk}=11.09(2)$ \cite{Wolffo3}.
 Hence the result from a lattice
of size $L=64$ should give $\xi_{bulk}$ to a good approximation.
The effective correlation length for the three different
 estimators at the distances
$\tau=10,20$ and 30 are given in table 5. The cluster-improved
 estimator is superior
to the new improved estimator but there is still
an advantage for the new improved
estimator compared to the conventional one by a factor of $1.4$ at $\tau=10$
that grows to $2.8$ at $\tau=30$ in the statistical error.

\subsection{The 2D $XY$ model. }

I simulated the 2D $XY$ model at $\beta=0.92$ in the massive phase, at
$\beta=1.1197$, which is the best estimate for the critical coupling
\cite{matching}
and at $\beta=1.2$ and $1.3$ in the massless phase. All simulations consist
of 25000 measurements.   The results for the correlation length obtained
from the different estimators are summarized
in table 6.

Let me again first discuss the reduction of the statistical error by the new
improved estimator.
 At $\beta=1.3$ the statistical error of the new estimator
is about 11 times smaller than  that of the conventional one and about 9 times
smaller than that of the cluster estimator.  At  $\beta=1.2 $ the gain in the
statistical accuracy
is about a factor of 9 compared to the conventional estimator and about 6
compared to the cluster estimator for lattices larger than $L=8$. At $\beta_c$
the gain is reduced to a factor of 6  and 4 respectively.
At $\beta=0.92$ the improvement depends very much on the lattice size.
While for the smallest lattice size $L=8$ there is still an improvement of
a factor of 2 compared to the cluster estimator, at $L=64$ the
cluster estimator is clearly better by a factor of almost 4.
On the other hand the new estimator stays more accurate than the
conventional estimator for all lattice sizes considered.

The dependence of the improvement on the distance $\tau$ is qualitatively
the same as for the $O(3)$ model. Therefore I omit a detailed discussion here.

Now let us turn to the physical interpretation of the results.
In the last column  of table 6 I give the running coupling
$\bar{g}^2$ computed from
the best estimate for $\xi$.
 For $\beta=1.2$ and $1.3$ the  values stabilize
for $L \ge 24$ within the small error bars.
 The small deviations for the smaller $L$ are consistent
 with corrections that die out with a negative power of $L$.
This behaviour nicely reveals the scale invariance of a massless theory.

For $\beta=\beta_c$ there is a tiny increase  of the coupling $\bar{g}^2$ over
the whole range of $L$ and the value for $L=64$ is still far away from the
predicted $L=\infty$ limit $\pi/2$. However this behaviour is completely
explained by
\be
\label{crit}
\bar{g}^2|_{\beta_c} =
 \frac{\pi}{2} - \frac{\pi}{4} \frac{g^2_0}{1+g^2_0 \ln L} \; ,
\ee
which can be derived from the KT-theory \cite{KT}.
Results of fits according eq. (\ref{crit})
 are
 given in  table 7. When lattices of size $L \le 8$ are discarded the fits
have an acceptable $\chi^2/dof$ and $g_0^2$ remains stable
when further
 data-points are discarded.

At $\beta=0.92$ $\bar{g}^2$ clearly increases with increasing lattice size $L$.
In order to measure $\xi_{bulk}$ one simulates lattices
with $L >> \xi_{bulk}$. Here  smaller distances than $\tau=L$ are sufficient
to get a good estimate for  $\xi_{bulk}$ from $\xi_{eff}$.
Therefore I evaluated $\xi_{eff}$ for $\beta=0.92$ and $L=64$ at various
distances. The results are given in table 8. At $\tau=10$ the statistical
error of the cluster estimator is much smaller than that of the new improved
estimator. But still the conventional one is worse by a factor of  1.3.

\section{Conclusion}
I showed that the new improved estimator drastically
reduces the statistical error
of the correlation length of finite systems. At values of the running
coupling $\bar{g}^2 $ of the 2D $O(3)$ model,
  such as used by L\"uscher et al. \cite{Luescher2}
the error was reduced by a factor of 4 up to  8 compared to the
cluster-improved estimator. At $\bar{g}^2 \approx 1/4$
a reduction of the error by a factor of 30 is reached.

On the other hand the error reduction seems to vanish in the limit
$ L/\xi_{bulk} \rightarrow \infty$. Nevertheless  at
$L \approx 6 \xi_{bulk}$, which might be sufficient for the determination
of the bulk correlation length, there is still an improvement of a factor
of  1.3 up to 1.4 compared to the conventional estimator,
 which might be interesting for models where no efficient cluster  algorithm
is known. A direct application of this method to gauge theories is not possible
since there is no exact global symmetry that could be employed for the
definition of an improved observable.
A possible extension of the idea is an implementation of a multigrid scheme
that has blocks with length one in time direction and 1,2,4 ... $L$ in spatial
direction.

As predicted, the running coupling $\bar{g}^2$ nicely signalled the presence
of the massless phase of the 2D $XY$ model. At fixed $\beta > \beta_c$
the coupling
 $\bar{g}^2$  rapidly approached a fixed point with increasing lattice size.

In contrast to that situation, at $\beta=5.0$ the $\bar{g}^2$ of the $O(3)$
model decreased with decreasing $L$. The change of $\bar{g}^2$  with doubling
the lattice size was more than 100 times the statistical error. The change of
 $\bar{g}^2$ with varying length scale is almost perfectly described by
the Callan-Symanzik $\beta$-function computed to
three-loop order in perturbation theory.

\section{Acknowledgement}
I would like to thank U. Wolff for many helpful discussions on the subject
of this article.  I thank K. Pinn and R. Sommer for a final discussion of
the manuscript.

\begin{table}[g]
\small
\begin{center}
\begin{tabular}{|c|c|r|l|l|l|l|}
\hline
\mc{1}{|c}{}  &
\mc{1}{|c}{$\beta$}  &
\mc{1}{|c}{$L$}  &
\mc{1}{|c}{$\xi_{con}$} &
\mc{1}{|c}{$\xi_{clu}$} &
\mc{1}{|c}{$\xi_{imp}$} &
\mc{1}{|c|} {$\bar{g}^2$} \\
\hline
 x &  1.9637 &  6 & \phantom{1}8.084(25)& \phantom{1}8.131(17)&
 \phantom{1}8.1295(25)&0.7381(2)\\
 x &  2.0100 &  8 &10.867(33)&10.835(29)&10.8288(30)&0.7388(2) \\
   &  2.0809 & 12 &16.339(58)&16.350(34)&16.2874(42)&0.7368(2)  \\
 x &  2.1260 & 16 &21.663(68)&21.715(43)&21.6645(50)&0.7385(2)   \\
   &  2.2408 & 32 &43.31(13) &43.321(82)&43.262(11) &0.7397(2)    \\
\hline
   & 1.9633 & 12 & 14.716(41)& 14.712(28)& 14.6969(42)&0.8165(2) \\
   & 2.0108 & 16 & 19.478(59)& 19.538(37)& 19.5873(51)&0.8169(2) \\
   & 2.0784 & 24 & 29.192(85)& 29.284(50)& 29.3862(75)&0.8167(2)  \\
   & 2.1264 & 32 & 39.17(11) & 39.288(71)& 39.1900(92)&0.8165(2) \\
   & 2.2428 & 64 & 78.34(22) & 78.50(14) & 78.450(20)& 0.8158(2)  \\
\hline
 x & 1.6050& 6 & \phantom{1}5.658(18)& \phantom{1}5.665(9) &
 \phantom{1}5.6651(22)&1.0591(4)\\
 x & 1.6589& 8 & \phantom{1}7.568(23)& \phantom{1}7.537(12)&
 \phantom{1}7.5571(30)&1.0586(4) \\
 x & 1.6982&10 & \phantom{1}9.433(29)& \phantom{1}9.422(14)&
  \phantom{1}9.4350(33)&1.0599(4)  \\
 x & 1.7306&12 &11.303(36)&11.333(17)&11.3247(41)&1.0596(4)   \\
 x & 1.7800&16 &15.075(40)&15.107(23)&15.1162(51)&1.0585(4)    \\
  & 1.8460&24 &22.597(66)&22.636(35)&22.6511(72)&1.0596(4)     \\
  & 1.8931&32 &30.041(91)&30.206(49)&30.1971(96)&1.0597(3)      \\
\hline
  & 1.6047& 12&\phantom{1}9.311(32)&\phantom{1}9.295(14)&
 \phantom{1}9.3040(43)&1.2898(6)  \\
  &1.6583& 16&12.462(41)  &12.458(20)&12.4548(59)  &1.2846(6)   \\
  &1.6985& 20&15.627(64)  &15.648(23)&15.6274(64)  &1.2798(5)    \\
  &1.7306& 24&18.760(60)  &18.849(28)&18.8174(77)  &1.2754(5)  \\
  &1.7793&32&25.148(82)  &25.193(33)&25.176(11) &1.2711(6)   \\
  &1.8460& 48&37.96(11) &37.927(49)&37.868(16) &1.2676(5)    \\
  &1.8932& 64&50.75(15) &50.694(70)&50.584(22) &1.2652(6)     \\
\hline
 \end{tabular}
  \parbox[t]{.85\textwidth}
  {
  \caption[XXXX
   ]{\label{tabene1}
Results for the correlation length $\xi$ of the 2D $O(3)$ model obtained from
various estimators of the correlation function. The subscripts
$con$, $clu$ and $imp$ denote the conventional, the cluster-improved and
the new improved estimator respectively. In addition I give the result for the
running coupling $\bar{g}^2$ obtained from the most accurate result for
$\xi$. The simulations that have a direct counter-part in table 2 of
ref. 4 are marked by an x.
   }
  }
 \end{center}
 \end{table}

\begin{table}[g]
\small
\begin{center}
\begin{tabular}{|c|c|c|}
\hline
$L/a$   &  $\delta_1$ & $\delta_2$ \\
\hline
 6      &  -0.00657 & 0.00289  \\
 8      &  -0.00364 & 0.00176  \\
 12     &  -0.00160 & 0.00085  \\
 16     &  -0.00090 & 0.00050  \\
\hline
 \end{tabular}
  \parbox[t]{.85\textwidth}
  {
  \caption[XXXX
   ]{\label{Luetab}
Perturbative results for the finite lattice spacing artefacts taken from
ref. \cite{Luescher2}.
   }
  }
 \end{center}
 \end{table}

 \begin{table}[g]
\small
\begin{center}
\begin{tabular}{|c|l|c|l|c|}
\hline
disc& $\sigma(2,1.0595)$   &   $c_2$  &\mc{1}{|c|}{$c_4 $}
 & $\chi^2/dof$\\
\hline
 0 & 1.2641(6) &  1.89(13)  & -34.7(3.0) & 0.38 \\
 1 & 1.2638(8) &  2.03(24)  & -44.(14.)  & 0.34 \\
 2 & 1.2639(10)&  1.95(47)  & -37.(42.)  & 0.49 \\
\hline
 \end{tabular}
  \parbox[t]{.85\textwidth}
  {
  \caption[XXXX
   ]{\label{tabene1}
Results of fits for $\Sigma(2,1.0595,a/L)$ according to eq. 38. disc gives
the number of discarded data points with small $L$.
   }
  }
 \end{center}
 \end{table}

\newpage

 \begin{table}[g]
\small
\begin{center}
\begin{tabular}{|r|c|c|c|c|c|c|}
\hline
 $L$  &    $\bar{g}^2$   & 1-loop& 2-loop& 3-loop&
$ \bar{g}^2 \delta(\bar{g}^2,a)$ \\
\hline
 6    &  0.22399(4)&0.22435&0.22373&0.22369 & -0.00030 \\
 8    &  0.22624(4)&0.22667&0.22613&0.22609 & -0.00017 \\
12    &  0.22959(4)&0.23004&0.22960&0.22957 & -0.00007 \\
16    &  0.23210(3)&0.23249&0.23213&0.23211 & -0.00004 \\
24    &  0.23570(3)&0.23603&0.23580&0.23579 &\\
32    &  0.23836(3)&0.23861&0.23847&0.23846 &\\
48    &  0.24234(3)&       &       &        &\\
\hline
 \end{tabular}
  \parbox[t]{.85\textwidth}
  {
  \caption[XXXX
   ]{\label{tabene1}
Results for the running coupling $\bar{g}^2$ at $\beta=5.0$ on lattices
of the size $L=6$ up to $48$. $\bar{g}^2$ is computed form the new improved
estimator of the correlation function. The Monte Carlo result is compared
with the result of the Callan-Symanzik $\beta$-function to
1, 2 and 3-loop order.
As starting point of the integration I took the MC result for $L=48$.
In the last column I give the finite lattice spacing artefact for a step
with scaling factor 2 as computed
from perturbation theory.
   }
  }
 \end{center}
 \end{table}

\begin{table}[g]
\small
\begin{center}
\begin{tabular}{|c|l|c|c|}
\hline
  $t$& \mc{1}{|c|}{$\xi_{con}$} & $\xi_{clu}$ & $\xi_{imp}$ \\
\hline
  11 & 11.023(33)  &  11.031(15) & 11.031(24)  \\
  22 & 10.923(85)  &  11.022(16) & 11.007(48)  \\
  33 & 10.70(26)   &  11.007(21) & 11.034(94)  \\
\hline
 \end{tabular}
  \parbox[t]{.85\textwidth}
  {
  \caption[XXXX
   ]{\label{tabene1}
Results for the effective correlation length for various
    distances $t$ at $L=64$ at $\beta=1.5$. The results can be
   compared with $\xi=11.09(2) $ obtained in ref. \cite{Wolffo3} on a $128^2$
   lattice.
   }
  }
 \end{center}
 \end{table}

\begin{table}[g]
\small
\begin{center}
\begin{tabular}{|c|r|l|l|l|l|}
\hline
\mc{1}{|c}{$\beta$}  &
\mc{1}{|c}{$L$}  &
\mc{1}{|c}{$\xi_{con}$} &
\mc{1}{|c}{$\xi_{clu}$} &
\mc{1}{|c}{$\xi_{imp}$} &
\mc{1}{|c|} {$\bar{g}^2$} \\
\hline
1.3 &   4   &\phantom{1}7.49(5)&\phantom{1}7.46(4)&\phantom{1}7.476(6)
 & \phantom{1}1.0701(9) \\
1.3 &   6   &  11.49(7)    &    11.44(5)     & 11.383(8)   &
\phantom{1}1.0542(7)  \\
1.3 &   8   &  15.26(10)   &    15.18(7)     & 15.265(10)  &
\phantom{1}1.0481(7)  \\
1.3 &  12   &  22.97(16)   &    22.93(12)    & 22.952(15)  &
\phantom{1}1.0457(7)  \\
1.3 &  16   &  30.65(21)   &    30.85(16)    & 30.661(18)  &
\phantom{1}1.0437(6)  \\
1.3 &  24   &  46.20(31)   &    46.15(23)    & 46.083(26)  &
\phantom{1}1.0416(6)  \\
1.3 &  32   &  61.25(39)   &    61.18(28)    & 61.420(35)  &
 \phantom{1}1.0420(6)  \\
1.3 &  48   &  92.25(65)   &    91.80(50)    & 92.192(49)  &
 \phantom{1}1.0413(6)  \\
\hline
1.2 &   4   &\phantom{1}6.53(4)&\phantom{1}6.54(3)&\phantom{1}6.502(7)&
 \phantom{1}1.2304(13) \\
1.2 &   6   &\phantom{1}9.74(6)&\phantom{1}9.75(4)&\phantom{1}9.856(9)&
\phantom{1}1.2175(11)  \\
1.2 &   8   & 13.19(9)     &   13.19(6)      & 13.189(12)  &
\phantom{1}1.2131(11) \\
1.2 &  12   & 20.05(14)    &   19.96(10)     & 19.839(16)  &
 \phantom{1}1.2097(10)  \\
1.2 &  16   & 26.43(17)    &   26.56(12)     & 26.419(22)  &
 \phantom{1}1.2112(10) \\
1.2 &  24   & 40.16(27)    &   39.59(18)     & 39.748(28)  &
 \phantom{1}1.2076(9)  \\
1.2 &  32   & 53.09(31)    &   53.07(22)     & 52.975(34)  &
 \phantom{1}1.2081(8)  \\
1.2 &  48   & 79.55(52)    &   79.59(40)     & 79.468(52)  &
 \phantom{1}1.2080(8) \\
\hline
1.1197&   4   &\phantom{1}5.75(4)&\phantom{1}5.74(3)&\phantom{1}5.695(7)&
\phantom{1}1.4047(17)  \\
1.1197&   6   &\phantom{1}8.57(6)&\phantom{1}8.57(3)&\phantom{1}8.554(10)&
\phantom{1}1.4029(16)   \\
1.1197&   8   &  11.39(7)    &   11.33(5)      & 11.333(13)  &
\phantom{1}1.4118(16)  \\
1.1197&  12   &  16.76(10)   &   16.87(7)      & 16.919(19)  &
\phantom{1}1.4185(16)  \\
1.1197&  16   &  22.45(15)   &   22.52(9)      & 22.446(23)  &
\phantom{1}1.4256(15)  \\
1.1197&  24   &  33.49(23)   &   33.38(14)     & 33.390(36)  &
\phantom{1}1.4376(15)  \\
1.1197&  32   &  44.26(29)   &   44.41(18)     & 44.369(42)  &
\phantom{1}1.4424(14)  \\
1.1197&  48   &  66.33(44)   &   65.92(30)     & 66.082(61)  &
\phantom{1}1.4527(13)  \\
1.1197&  64   &  88.59(59)   &   87.39(35)     & 87.898(89)  &
\phantom{1}1.4562(15)  \\
$\beta_c$& $\infty$&         &                 &             &
\phantom{1}1.570796  \\
\hline
0.92 &  4   &\phantom{1}3.73(3)&\phantom{1}3.706(12)&\phantom{1}3.688(6)&
\phantom{1}2.169(4)  \\
0.92 &  8   &\phantom{1}6.00(6)&\phantom{1}6.044(18)&\phantom{1}6.018(11)&
\phantom{1}2.659(3)\\
0.92 & 16   &\phantom{1}8.46(9)&\phantom{1}8.578(25)&\phantom{1}8.630(25)&
\phantom{1}3.708(11)\\
0.92 & 32   &  10.78(28)   & 10.193(26)      &10.221(86)   &
 \phantom{1}6.279(16)  \\
0.92 & 64   &              & 10.559(42)      &
 \phantom{1}8.8(1.2)    &12.12(5)   \\
\hline
 \end{tabular}
  \parbox[t]{.85\textwidth}
  {
  \caption[XXXX
   ]{\label{tabene1}
Results for the correlation length $\xi$ of the 2D $XY$ model obtained from
different estimators of the correlation function. The subscripts
$con$, $clu$ and $imp$ denote the conventional, the cluster-improved and
the new improved estimator respectively. In addition I give the result for the
running coupling $\bar{g}^2$ obtained from the most accurate result for
$\xi$.
   }
  }
 \end{center}
 \end{table}

\begin{table}[g]
\small
\begin{center}
\begin{tabular}{|c|c|c|c|c|c|}
\hline
$disc$       &    0     &   1      &   2     &  3      &   4     \\
$g_0^2$     & 0.3466(4)& 0.3618(4)& 0.367(5)& 0.372(6)&  0.372(7)\\
$\chi^2/dof$&  8.4      &  1.4     & 0.9     &  0.2    & 0.3     \\
\hline
 \end{tabular}
  \parbox[t]{.85\textwidth}
  {
  \caption[XXXX
   ]{\label{tabene1}
 Results for fits of the running coupling $\bar{g}^2$ of the 2D $XY$
 model at $\beta_c $ following eq. 39. $disc$ gives the number of data-points
 with small $L$ that are discarded.
   }
  }
 \end{center}
 \end{table}

\begin{table}[g]
\small
\begin{center}
\begin{tabular}{|c|l|l|l|}
\hline
  $\tau$ & \mc{1}{|c|}{$\xi_{con}$} &
 \mc{1}{|c|}{$\xi_{clu}$} &\mc{1}{|c|}{$\xi_{imp}$} \\
\hline
   10 & 10.568(24)  &   10.552(8) & 10.568(18) \\
   20 & 10.56(6)    &   10.556(11)& 10.53(4)  \\
   30 & 10.66(17)   &   10.537(14)& 10.55(9)  \\
\hline
\end{tabular}
  \parbox[t]{.85\textwidth}
  {
  \caption[XXXX
   ]{\label{tabene1}
Results for the correlation length of the 2D $XY$ model at $\beta=0.92$ and
$L=64$. The effective correlation length is computed from the
conventional $con$, the new improved $imp$ and the cluster-improved $clu$
estimator at the distances $\tau=10,20$ and $30$. The results can be
compared with $\xi=10.69(8) $  obtained form a $128^2$ lattice \cite{Wolffxy}.
   }
  }
 \end{center}
 \end{table}

\newpage
{\large \bf Figure captions}

\vspace{2cm}
Figure 1:
 The relative error of the time-slice
 correlation function $\langle G(\tau) \rangle$ computed from the
 conventional, the cluster-improved and the new improved estimator
 as a function of the distance $\tau$ for $L=64$ and $\beta=1.8932$.
\vspace{2cm}

 Figure 2:
 The relative error of the effective
 correlation length
 computed from the
 conventional, the cluster-improved and the new improved estimator
 of the correlation function
 as a function of the distance $\tau$ for $L=64$ and $\beta=1.8932$.

\end{document}